\documentclass{optica-article}

\journal{opticajournal} 

\articletype{Research Article}

\usepackage{lineno}
\usepackage{xcolor}
\usepackage{comment}
\usepackage{siunitx}
\usepackage{graphicx}

 \usepackage{todonotes}

\begin{document}

\title{Amplified Feedback and Spontaneous Emission Injection in Quantum Cascade Ring Laser Systems}

\author{ Radhika R. Bhuckory,\authormark{1,*} Sara Kacmoli,\authormark{1} Deborah L. Sivco,\authormark{1,2} and Claire Gmachl\authormark{1}}

\address{\authormark{1}Department of Electrical and Computer Engineering, Princeton University, Princeton, New Jersey 08544, USA}

\address{\authormark{2}Currently with Trumpf Photonics Inc., 2601 US Highway 130, Cranbury, NJ 08512, USA}

\email{\authormark{*}rb4130@princeton.edu} 

\begin{abstract*} Ring lasers exhibit rich operational regimes such as unidirectional, bidirectional, or bistable operation. The two-mode dynamics of the counter-propagating modes - clockwise (CW) and counterclockwise (CCW) and their selection, have gained attention in the mid-infrared; however, the underlying switching mechanism has been largely unexplored. Previously, we experimentally demonstrated robust and deterministic mode selection in a ring quantum cascade laser (QCL) with an active outcoupling waveguide. Here, we apply the Lang-Kobayashi framework to our system to model the effects of spontaneous emission in the waveguide arms as well as amplified optical feedback from the facets on the mechanics of mode switching. We find that coherent feedback from facet reflections agrees upwards of 93\% with the experimental behavior, indicating that amplified feedback is the dominant mechanism driving the mode selection dynamics.

\end{abstract*}

\section{Introduction}

Ring lasers and their dynamics have been thoroughly studied in various material platforms ~\cite{sorel2003ringlasers,Bradley2010,Luo2021,Ma2025} and continue to be at the forefront of theoretical and experimental studies ~\cite{sorel2002alternate,akparov2010ringlaser, meng2020mid,Kazakov2024}. The two-mode dynamics in ring laser systems are largely governed by the nonlinear interaction of the clockwise (CW) and counterclockwise (CCW) fields. In an ideal resonator, these two spatial modes do not couple to each other, giving rise to the traveling-wave nature of the total electric field that evade the effects of spatial hole burning (SHB). The two counterpropagating modes give rise to rich dynamical regimes such as unidirectional and bidirectional operation, bistable operation, and alternate oscillations. These regimes have been demonstrated in experiment ~\cite{sorel2002unidirectional,perez2007bistability} and treated in theory via the Lang-Kobayashi framework ~\cite{lang1980external}.

Recently, ring lasers have attracted attention in the mid-infrared due to the implementation of ring quantum cascade laser (QCL) systems and coherent instabilities therein ~\cite{meng2020mid,Heckelmann2023,Kazakov2024}. The QCL active medium is ideal for coupling light into in-plane resonator geometries ~\cite{kacmoli2024qclring}. The unidirectional operation exhibited by ring QCLs has been shown to be necessary for realizing active mode locking, coherent ring-based frequency comb generation, and high-sensitivity spectroscopic detection ~\cite{meng2020mid,malara2013ringqcl,Heckelmann2023}.

Our previous studies have included the fabrication and characterization of a unidirectional monolithic racetrack QCL~\cite{kacmoli2022unidirectional}. We demonstrated deterministic and controllable selection between the two counterpropagating racetrack modes from the differentially pumped waveguide arms. A small current differential can trigger a full and permanent switch between the CW and CCW modes. However, little is known about the mechanism behind the mode switching that we observe. 

In this work, we model the complex dynamics of a monolithic racetrack QCL to identify the dominant mechanism behind the observed switching between the CW and CCW modes observed in experiment ~\cite{kacmoli2022unidirectional}. We apply and adapt the Lang–Kobayashi framework \cite{lang1980external} to our QC ring laser system, which are deterministic rate equations used to describe a semiconductor laser with external optical feedback. We investigate two possible causes for the mode selection we observe: injection of spontaneous emission from the coupled waveguide arm, or amplified feedback from the facet of the outcoupling waveguide. Results from the simulation show strong agreement between the latter explanation and experimental data, suggesting that coherent feedback is the dominant mechanism behind mode selection and switching in our QC ring laser system.

\section{Device Characteristics}

\begin{figure}[htbp]
\centering\includegraphics[scale=0.45]{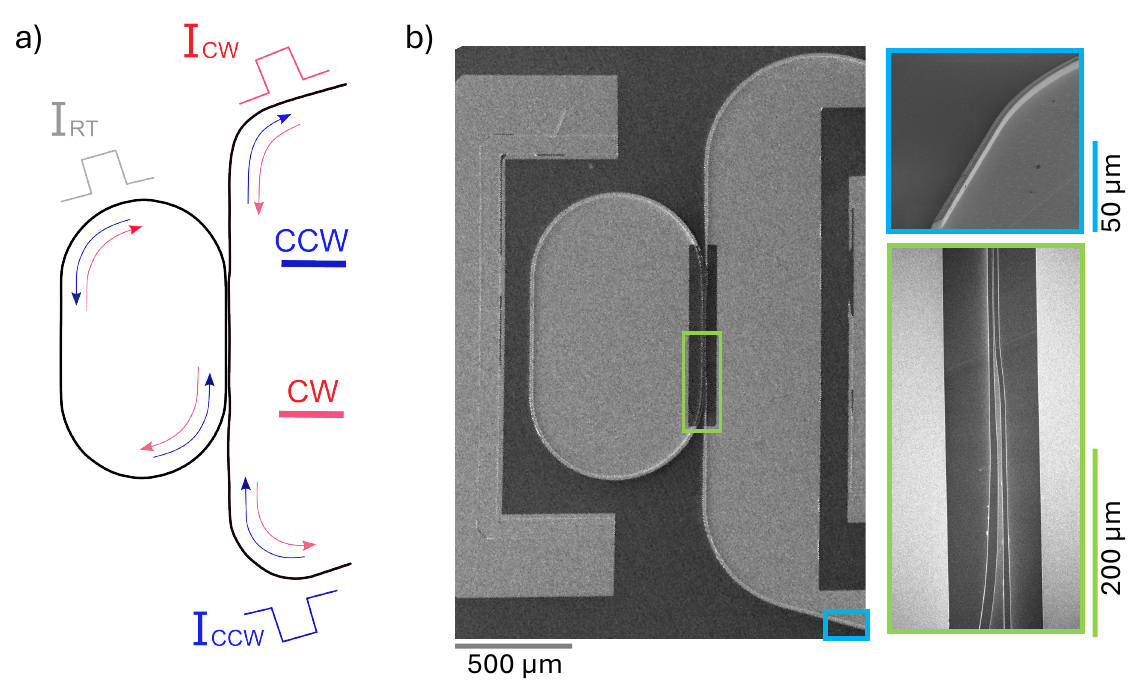}
\caption{ (a) Device schematic showing the racetrack laser (left) and the evanescently-coupled waveguide (right). The blue and red arrows show the direction of clockwise (CW) and counter-clockwise (CCW) modes. $I_{\text{RT}}$, $I_{\text{CW}}$, $I_{\text{CCW}}$ are the current pulses to the racetrack and waveguide arms. (b) Top view scanning electron microscope (SEM) image of a similar device detailing (top panel) left waveguide arm angled at \(17^\circ\) relative to the cleaved facet; and (bottom panel) a top view of the taper geometry of the coupler.}
\label{schematic}
\end{figure}

In Ref.~\cite{kacmoli2022unidirectional} we report a monolithic racetrack QCL that is fabricated on a standard $\lambda \approx 8.8\,\mu$m QC wafer based on a double phonon resonance design ~\cite{Liu2006}. The racetrack resonator is evanescently coupled to an active bus waveguide on the same chip, leading to intersubband transitions at the same wavelength emitted by the racetrack. This also enables the outcoupling waveguides to be pumped to transparency to overcome absorption. A schematic of the design and a top-view scanning electron microscope (SEM) image of a simialr fabricated device are shown in Fig.~\ref{schematic}.


The key features of the system are the racetrack, waveguide arms, and coupling region. The racetrack and waveguide arms each have their own metal contact and ground pads, allowing for each component to be controlled independently. To minimize asymmetry in their optical properties, both waveguide arms are designed to have equal path lengths. 


The coupling region plays a critical role in the system’s operation. The width of the coupling region is selected as the minimum width required to support the fundamental mode, thereby enhancing the evanescent coupling. The fabrication processes developed in our prior work are optimized for low sidewall roughness and vertical sidewalls. This in turn reduces absorption and back-scattering, which limits the coupling between the counter-propagating modes.

The all-active nature of the waveguide introduces two key dynamics: spontaneous emission (SE) and amplified feedback (AF). The bus waveguide generates its own spontaneous emission, which, depending on the pumping level, is amplified by the active media and evanescently coupled into the racetrack. Additionally, the waveguide arms are angled at \(17^\circ\) relative to the cleaved facet as shown in Fig.~\ref{schematic}b, resulting in a reflectivity of approximately 1\% \cite{Ahn2013,Aung2014}. This angle of the waveguide arms is chosen to minimize feedback. However, since the waveguide is active, any small amount of feedback that is reflected back into the system can be amplified depending on the pumping level. The unique contribution that each mechanism has on the switching dynamics or which mechanism is dominant is studied here.

\section{Mode selection dynamics}

Prior work has shown that the ring laser randomly selects between the CW or CCW modes with equal probability. By applying a current differential, $\Delta I$, to the two waveguide arms, we can deterministically select one of the counter-propagating modes of the racetrack laser ~\cite{kacmoli2022unidirectional}.

To examine the statistics of switching between these modes, a current, $I_0 \pm \Delta I $, is applied to the waveguide arms to measure the amount of current differential, $\Delta I$, that is needed to fully switch the mode. By repeating this measurement 15 times per average current, $I_0$, we obtain a probability that the CW (CCW) mode dominates under a specific applied current configuration. A representative set of individual traces observed by the detector at one of the waveguide arms, used to compute the probability, is shown in Fig.~2a. We note that our racetrack laser is fully unidirectional, i.e., we never observe a signal on both the CW and CCW photodetector at the same time.

\begin{figure}[htbp]
\centering\includegraphics[scale=0.6]{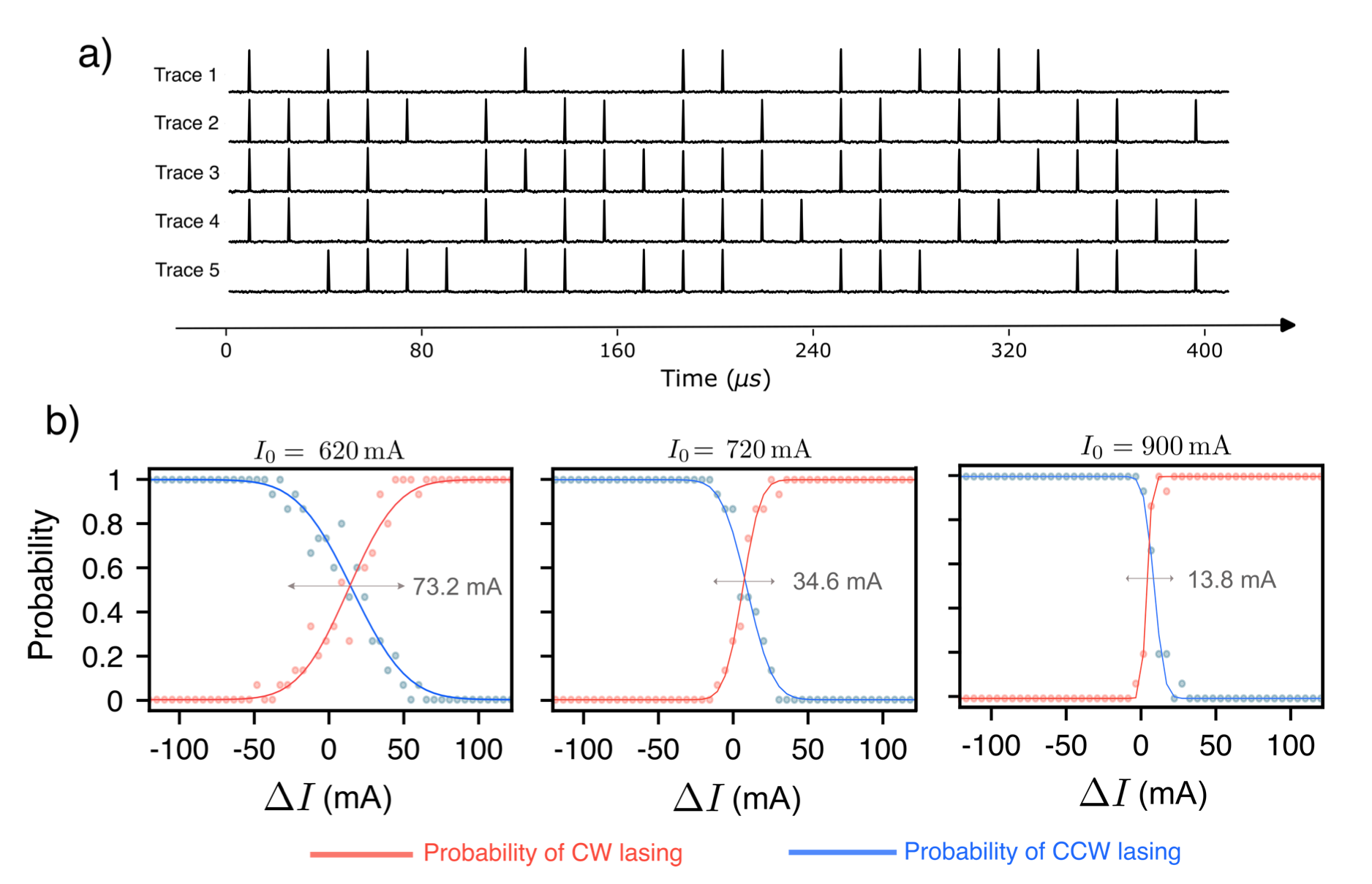}
\caption{(a) Five representative traces measured at one of the waveguide facets, illustrating the received pulse trains used to obtain the probability of each mode dominating. The traces demonstrate the random selection between CW and CCW modes in the absence of a bias applied to the waveguide arms. (b) Probability that the mode is CW (red dots) or CCW (blue dots) from 15 consecutive measurements for three different values of the average input current, $I_0$. Solid lines represent an error function fit. The input current imbalance, $\Delta I$, from 10\%-90\% probability, needed for mode switching decreases with increasing $I_0$.}
\end{figure}

The results of this experiment are shown in Fig.~2b for three different values of $I_0$. In red (blue), we show the probability that the CW (CCW) mode dominates. Note the crossover happens at $\Delta I = 0\,\mathrm{mA}$. Another key detail is the span of the bistable region (where probabilities of each mode dominating are clearly less than 100\%). To determine this span, we fit our data using the error function and calculate the width going from 10\%--90\% probability. For $I_0 = 620\,\mathrm{mA}$ this region spans $\sim 73.2\,\mathrm{mA}$. As $I_0$ increases to $900\,\mathrm{mA}$, the span of this region decreases to 13.8 mA.

\section{The mechanics of switching: amplified feedback and spontaneous emission}

Thus far we have shown experimental results from our previous studies that describe the key characteristics of these active waveguide couplers -- the deterministic mode selection of the rotational modes of the ring laser. Now, we turn to studying these devices in theory in order to obtain insight into the physics behind the switching between the two modes that we have observed in experiment. 

A simplified model often used to simulate semiconductor laser dynamics is given by the Lamb equations which take the following form:

\begin{equation}
\dot{E}_{1,2} = \frac{1 + i\alpha}{2} 
\left[ N\left(1 - s|E_{1,2}|^2 - c|E_{2,1}|^2\right) - 1 \right]E_{1,2} - 
\left(k_d + i k_c\right)E_{2,1}
\label{eq1}
\end{equation}

\begin{equation}
\dot{N} = \gamma\left[\mu - N - N\left(1 - s|E_1|^2 - c|E_2|^2\right)|E_1|^2 
- N\left(1 - s|E_2|^2 - c|E_1|^2\right)|E_2|^2 \right]
\label{eq2}
\end{equation}

The Lamb equations describe the two slowly varying complex amplitudes of the counterpropagating fields: $E_1$ (CW mode) and $E_2$ (CCW mode) \cite{perez2007bistability}. In these equations, $\alpha$ represents the linewidth enhancement factor or amplitude-phase coupling, which is relatively small (0.5--2) for QC lasers \cite{Jumpertz2016}. The most significant dynamics in the system are captured by parameters $s$ and $c$, where $s$ corresponds to gain self-saturation and $c$ corresponds to gain cross-saturation which is one of the terms that nonlinearly couples the two counterpropagating fields \cite{perez2007bistability}. Another coupling term arises from backscattering and is characterized by two components: $k_d$, the dissipative backscattering term accounting for localized losses and absorption, and $k_c$, the conservative backscattering term representing scattering and reflection \cite{lenstra1990rings}. In semiconductor lasers, conservative backscattering typically dominates because of reflections caused by imperfections introduced during fabrication, leading to standing wave patterns inside the ring cavity. Additionally, the evanescent coupler itself serves as a source of reflections. Finally, $\gamma$ is the ratio of the photon to carrier lifetime, which is much greater in QC lasers compared to other semiconductor lasers due to the ultrafast (on the order of $\sim$ps) carrier lifetime \cite{Faist2022}. The parameter $\mu$ denotes the pump power applied to the system. The pump power, $\mu$, is normalized by the lasing pump threshold. Polarization dynamics, typically present in laser models, are adiabatically eliminated in this formulation. Eq. 2, characterizes the dimensionless carrier density, $N$. In this model, it is evident that the two equations for the counterpropagating modes are fully symmetric. If the system operates in a stable regime, the winning mode is determined randomly. To validate the results of our implementation, we simulate these equations over multiple iterations and indeed find that the probability of clockwise or counterclockwise mode dominance is 50\% (not shown).

Next, we investigate two additional terms that can introduce asymmetry into these equations, consistent with the experimental observations and our inclusion of the bus waveguide. With these terms included, the system is governed by the Lang-Kobayashi equations \cite{lang1980external}:

\begin{align}
\dot{E}_{1,2} &= 
\frac{1 + i\alpha}{2} 
\left[ N \left( 1 - s|E_{1,2}|^2 - c|E_{2,1}|^2 \right) - 1 \right]E_{1,2} 
- (k_d + i k_c)E_{2,1} 
+ \sqrt{D_{1,2}} \xi_{}
+ F_{1,2}E_{2,1}(t - \tau), \label{eq:3} \\
\dot{N} &= \gamma \bigg[ \mu - N - N\left(1 - s|E_1|^2 - c|E_2|^2\right)|E_1|^2 
- N\left(1 - s|E_2|^2 - c|E_1|^2\right)|E_2|^2 \bigg]. \label{eq:4}
\end{align}

The first added term represents the spontaneous emission, where \( D_{1,2} \) is the strength of the spontaneous emission in the ring, formally given as 

\[D_{1,2} = C_{s1,2}(N + G_0 \tau_p N_0).\]

In this expression, \( C_{s1,2} \) is the spontaneous emission factor, \( G_0 \) is the differential gain, and \( N_0 \) is the transparent carrier density. \( \xi \) represents an independent complex white Gaussian noise with zero mean and unit variance. The white noise term on the right-hand side turns this set of equations into stochastic differential equations (SDEs), where the random value updates with every integration step.

The second term added to the Lamb Equations in Eqs. 1 and 2 represent the presence of feedback, the true signature of the Lang-Kobayashi equations. \( F_{1,2} \) is the strength of feedback, which is generally a complex quantity. The amplitude is given by the expression 

\[
F_{1,2} = f_{1,2} \cdot b^2 \cdot e^{i\phi},
\] accounting for the two instances where light is coupled out of the ring and back in, represented by the term \( b^2 \). While the \( 17^\circ \) angle of the facet provides a low reflectivity, \( r \), of only \( \sim 1.1\% \), once current is applied to the waveguide arms, the reflected light is amplified. Therefore, the term \( f_{1,2} \) accounts for the gain incurred on each pass, as well as the reflection from the facet, the term which we generally refer to as amplified feedback. Feedback couples the two counterpropagating modes in a linear fashion; however, a delay \( \tau \) is necessary to account for the roundtrip time from the facet. To simplify simulations, we assume that the feedback delay is instantaneous by setting \( \tau \) to zero. Because the amplified feedback term is a complex term, it also accounts for a phase shift due to the reflection from the facets.

A summary of the coefficients and their corresponding values used in the numerical simulations are presented in Table 1. These values were chosen to best model a ring QCL and also reflect experimental parameters wherever applicable.

\begin{table}[h!]
    \centering
    \renewcommand{\arraystretch}{1.2} 
    \setlength{\tabcolsep}{4pt} 
    \small 
    \begin{tabular}{c|l|c|c|}
        \hline
        \multirow{9}{*}{\rotatebox{90}{\parbox{2.5cm}{\centering \textbf{Lamb Model}}}} 
        & \textbf{Symbol} & \textbf{Quantity} & \textbf{Typical values for ring QCLs} \\ \cline{2-4}
        & $\alpha$ & Linewidth enhancement factor & 0.5 \cite{Jumpertz2016}
 \\ \cline{2-4}
        & $s$ & Gain self-saturation & 0.2 \cite{Faist2022} \\ \cline{2-4}
        & $c$ & Gain cross-saturation & 0.4 \cite{Meystre1999} \\ \cline{2-4}
        & $k_c$ & Conservative backscattering coefficient & 0.0078 \cite{Haus1985} \\ \cline{2-4}
        & $k_d$ & Dissipative backscattering coefficient & 0.0003 \cite{Haus1985} \\ \cline{2-4}
        & $\gamma$ & Ratio of photon to carrier lifetime & 40 \cite{Faist2022} \\ \cline{2-4}
        & $\mu$ & Normalized pump power & 1.45 \\ \hline
        \multirow{6}{*}{\rotatebox{90}{\parbox{2.5cm}{\centering \textbf{Spontaneous Emission}}}} 
        & $C_{s1}$ & Spontaneous emission factor & $5 \times 10^{-8}$ \\ \cline{2-4}
        & $C_{s2}$ & Spontaneous emission factor & $5 \times 10^{-8}$ \\ \cline{2-4}
        & $G_0$ & Differential gain & $10 \times 10^{-12}$ m$^3 \cdot$s$^{-1}$ \cite{Vukovic2016,Spitz2021chaos} \\ \cline{2-4}
        & $\tau_{p}$ & Photon lifetime & 40 ps \cite{Faist2022}  \\ \cline{2-4}
        & $N_0$ & Transparent carrier density & $1.4 \times 10^{-24}$ m$^{-3}$ \cite{Vukovic2016} \\ \hline
        \multirow{6}{*}{\rotatebox{90}{\parbox{2.5cm}{\centering \textbf{Amplified Feedback}}}} 
        & $\text{R}_1$ & Reflection coefficient (right arm) & 0.01 \\ \cline{2-4}
        & $\text{R}_2$ & Reflection coefficient (left arm) & 0.01 \\ \cline{2-4}
        & $\text{b}$ & Coupling coefficient & 0.2 \\ \cline{2-4}
        & $\phi_1$ & Phase delay (right arm) & $\pi$ \\ \cline{2-4}
        & $\phi_2$ & Phase delay (left arm) & $\pi$ \\ \hline
    \end{tabular}
   \caption{Summary of symbols present in the Lang-Kobayashi model and typical values used to model a ring QCL, categorized by contributions from the Lamb model, and the two terms representing Spontaneous Emission and Amplified Feedback.}
    \label{tab:coefficients}
\end{table}

\section{Numerical Simulations}
\subsection{Amplified Feedback}

To mirror the experimental results shown in Fig.~2, we set up the simulations with an asymmetry between the waveguide arms. First, this asymmetry is applied as a difference in amplified feedback, \( \Delta f \), between the clockwise and counterclockwise modes. To ensure that a steady state is reached, we run the simulations for 1000 round-trips of the bus waveguide. For each value of the amplified feedback, 16 simulations are performed. We compute the probability that each mode dominates across the 16 trials for every difference in amplified feedback and run this entire simulation for various values of the pump power \( \mu \). The results of these simulations for \( \mu = 1.45\) that closely match the experimental pump power are shown in Fig.~3. 

We fit the data using an error function, similar to the experimental data, and calculate the width of the probability interval 10\% to 90\%. Using this calculated width, \( \Delta f \), and the average feedback value, \( f_0 \), we create a metric \( \Delta f / f_0 \) to compare with the experimental analogous ratio \( \Delta I / I_0 \). However, these two ratios are not fully equivalent because, in the case of feedback, light traverses the waveguide arm and experiences gain twice. To account for this, a square root factor is introduced. Fig. 3 show the three \( \Delta f / f_0 \) ratios that best match the experimental results.

\begin{figure}[htbp]
\centering\includegraphics[scale=0.40]{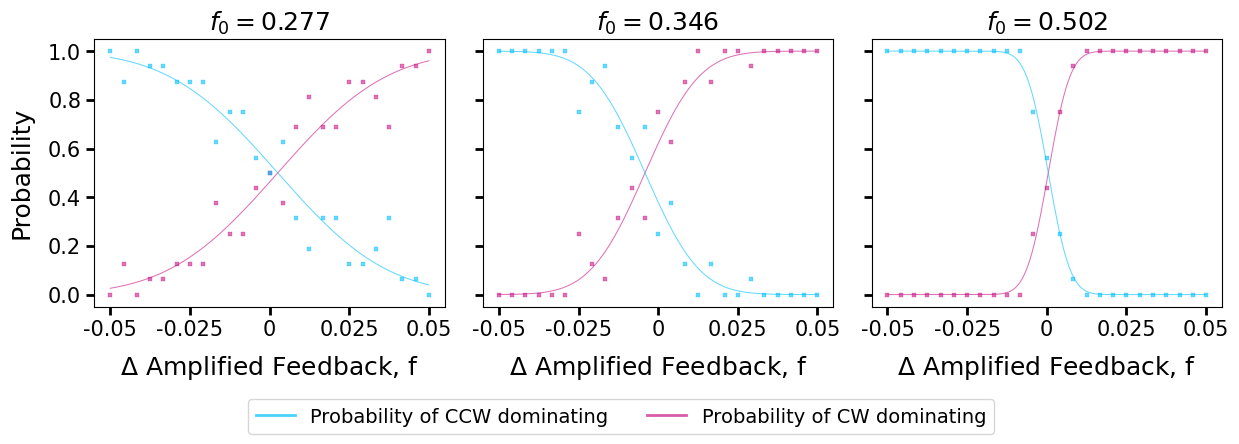}
\caption{Computed probability of each mode dominating as a function of the imbalance in amplified feedback coupled into the CW and CCW modes for three values of average amplified feedback, $f_0$, as computed by our simulations. Dots depict the data and solid lines are the error function fits of each data set. The parameters used in this simulation are $\alpha = 0.5$, $s = 0.2$, $c = 0.4$, $k_c = 0.0078$, $k_d = 0.0003$, $\gamma = 40$, $b = 0.2$, and $\varphi = \pi$. The pump power $\mu = 1.45$ is normalized with experimental current density threshold, $J_{th}$.}
\end{figure}

To quantify the goodness-of-fit of the amplified feedback (AF) data to the error function curve, we compute various statistical metrics such as the root mean squared error (RMSE) and the coefficient of determination (\( R^2 \)) and report them for each $f_0$ in Table~\ref{tab:fit_metrics}.

\begin{table}[h]
    \centering
    \small 
    \renewcommand{\arraystretch}{1.1}
    \begin{tabular}{c|cc|cc|cc}
        \hline
        \multirow{2}{*}{Metric} & \multicolumn{2}{c|}{$f_0 = 0.277$} & \multicolumn{2}{c|}{$f_0 = 0.346$} & \multicolumn{2}{c}{$f_0 = 0.502$} \\
        & E1 & E2 & E1 & E2 & E1 & E2 \\
        \hline
        RMSE          & 0.0833  & 0.0833  & 0.0753  & 0.0753  & 0.0165  & 0.0165  \\
        $R^2$ Score   & 0.9389  & 0.9389  & 0.9683  & 0.9683  & 0.9988  & 0.9988  \\
        \hline
    \end{tabular}
    \caption{Goodness-of-fit metrics for Amplified Feedback data fitted to an error function curve at different $f_0$ values.}
    \label{tab:fit_metrics}
\end{table}

The statistical quantities presented in Table~\ref{tab:fit_metrics} demonstrate that upwards of 93\% of the data follows the error function model across different values of $f_0$. At $f_0 = 0.502$, the $R^2$ value reaches 0.9988, indicating an almost perfect fit. The root mean square error (RMSE) decreases by approximately 80\% from 0.0833 at $f_0 = 0.277$ to 0.0165 at $f_0 = 0.502$, showing that the residual differences between the data and the fit decrease as $f_0$ increases. This suggests that the error function more accurately captures the transition between either of the modes dominating at higher $f_0$ values. The results from the simulation reveal that the underlying transition dynamics of the system are well captured by the error function, particularly at larger $f_0$ values similar to experiment.

The sensitivity to switching can be characterized by \( \Delta f \), the width of the bistable region, where the probabilities of each mode dominating clearly less than 100\%. To more systematically analyze the effects of the average feedback value, $f_0$, we perform the same simulations for 15 different $f_0$ values, calculating the width of the bistable region and plotting it as a function of $f_0$. We compute the statistical metrics for each data set and highlight widths where the $R^2$ value falls below 90\% in red, as shown in Figure~4. The results indicate that as the average feedback value increases, the system exhibits increased sensitivity of the switch due to more light being reflected back into the system, consistent with our experimental findings.

\begin{figure}[htbp]
\centering\includegraphics[scale=0.5]{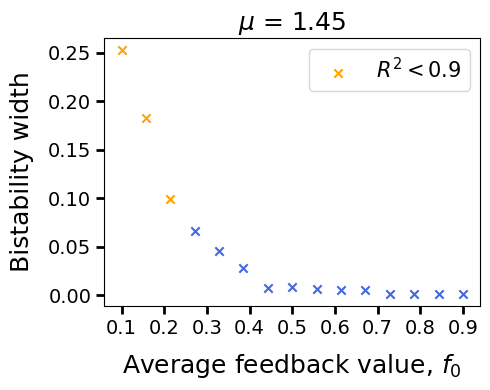}
 \caption{Impact of average feedback $f_0$ on the bistability width (10\%–90\% probability) of the error function fit at a fixed pump power $\mu = 1.45$. Data points in orange indicate error function fits with $R^2 < 0.9$.}
\end{figure}

\subsection{Spontaneous Emission Injection}

Second, we set up simulations where the asymmetry between the two waveguide arms is determined by the difference in the strength of the spontaneous emission (SE) injection, $C_{s}$ while the feedback is turned off. The results of these simulations are shown in Fig.~5 for three different values of average SE strength accross three orders of magnitude. An error function fit is not appropriate for these data sets, indicating that SE injection is not the primary mechanism controlling the mechanics the system. 

\begin{figure}[htbp]
\includegraphics[scale=0.43]{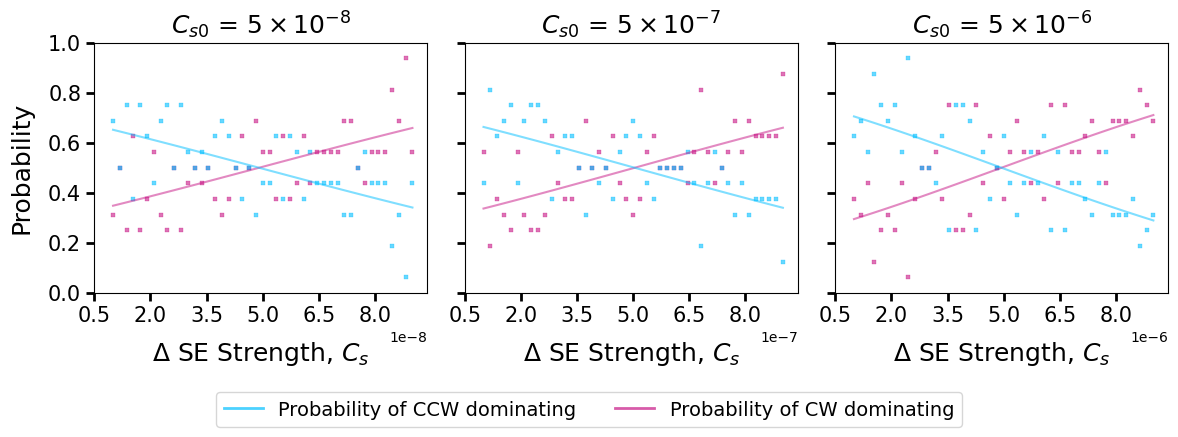}
\caption{Probability of each mode dominating as a function of the imbalance in the spontaneous emission strength coupled into the CW and CCW modes for three values of average spontaneous emission strength across three orders of magnitude. Dots depict the data and solid lines are the optimal error function fits of each data set. The parameters used in this simulation are $\alpha = 0.5$, $s = 0.2$, $c = 0.4$, $k_c = 0.0078$, $k_d = 0.0003$, $\gamma = 40$. The pump power $\mu = 1.45$ is normalized with experimental current density threshold, $J_{th}$. Additionally, $\tau_p = 40$ ps, $G_0 = 10 \times 10^{-12}$ m$^3$/s, and $N_0 = 1.4 \times 10^{-24}$ m$^{-3}$.}
\end{figure}

 Although Fig.~5 visually demonstrates that an error function fit does not capture the mode switching dynamics well, we compute the same statistical metrics for completeness. While there is some imbalance resulting from SE injection between the CW and CCW modes, the statistical metrics confirm that amplified feedback (AF) follows the error function model significantly better than spontaneous emission (SE) with more than 93\% of the data being captured by the fit for AF compared to only 47\% for SE for \( C_s = 5 \times 10^{-6} \). This means that at its best, SE captures less than half of the variance in the data. The results indicate that SE may contribute to mode selection and switching; however, this effect is more muted and qualitatively different from what we observe experimentally.

 \begin{figure}[htbp]
\centering\includegraphics[scale=0.47]{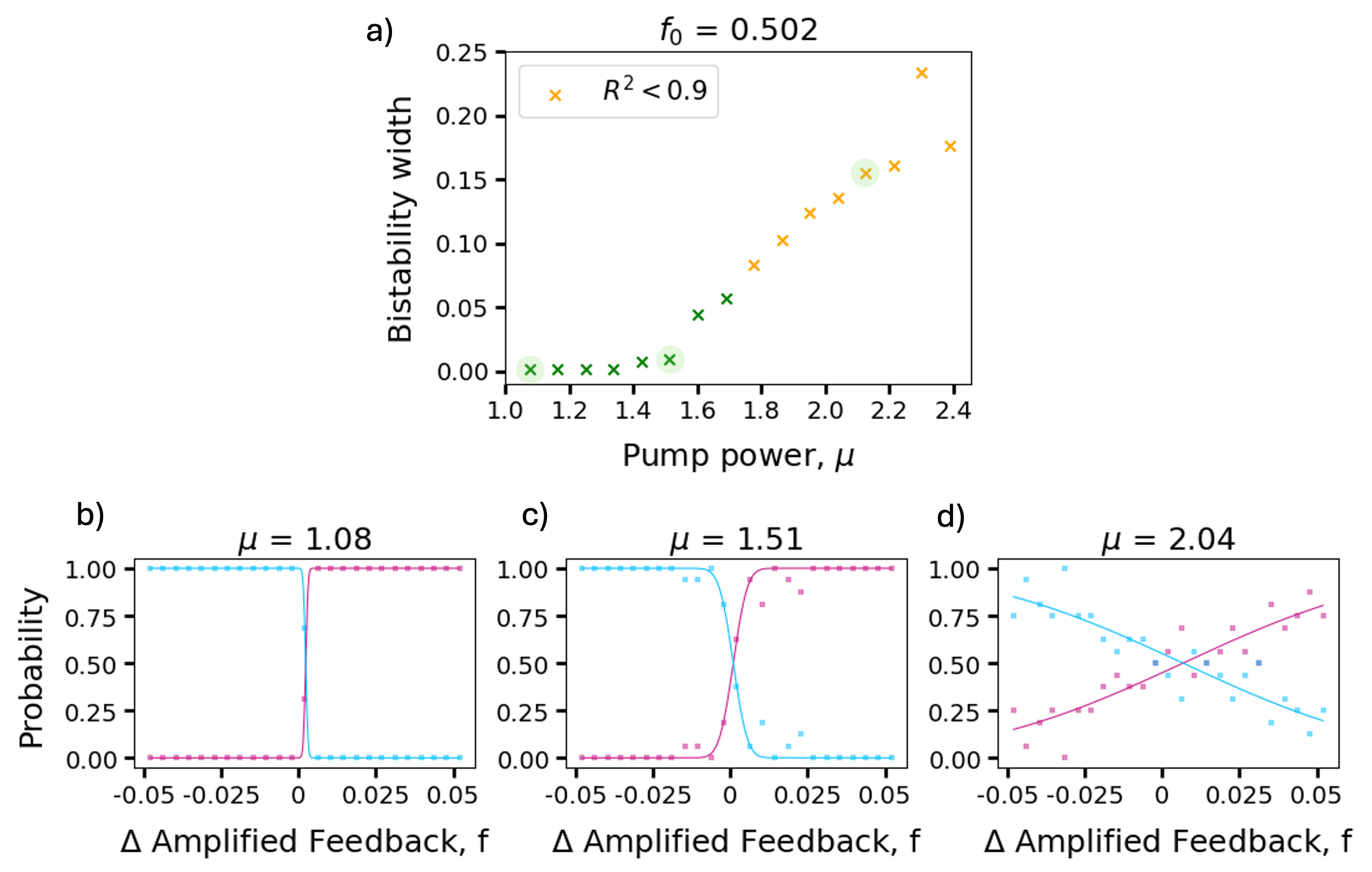}
\caption{(a) Impact of the pump power, $\mu$, normalized with the experimental current density threshold ( $J_{th}$), on the bistability width (10\%–90\% probability) for $f_0 = 0.502$. Data points in orange indicate error function fits with $R^2 < 0.9$. (b–d) Computed probability of each mode dominating as a function of the imbalance in amplified feedback for three values of $\mu$ highlighted in green in a), with an error function fit applied to each data set.}

\end{figure}

\subsection{Pump Power}

All the simulation results presented thus far have been for a fixed pump power of $\mu = 1.45$, corresponding to the current injection applied to the racetrack normalized by the lasing threshold, $J_{th}$. To further explore the influence of amplified feedback (AF) on the system, we perform a parameter sweep over multiple pump power values (\(\mu\)) for a fixed average feedback value of \( f_0 = 0.502 \). For each simulation, we compute the width of the bistable region along with the corresponding goodness-of-fit metrics and present the results in Fig.~6. 
As these simulation results suggest, it is plausible that at higher values of pump power, \( \mu \), the system exhibits lower sensitivity to feedback than at lower values. However, this interpretation requires experimental validation for future work.

\section{Conclusion}

In this paper, we have applied the Lamb-Kobayashi framework to a single ring QCL coupled to an active bus waveguide to simulate the effects of coherent feedback and incoherent light injection on the mechanics of mode switching -- amplified feedback at the facets and spontaneous emission in the waveguide arms, respectively. The results of the numerical simulations demonstrate amplified feedback to be the dominant mechanism driving the mode selection dynamics. We use an error function to fit the experimental and numerical results. The increased sensitivity of a switch between the CW and CCW modes to an increase in feedback is consistent with the behavior of the switching mechanism that we observe in experiments ~\cite{kacmoli2022unidirectional}. The results from modeling the SE term in the framework indicate that SE, that is amplified in the waveguide arms, may contribute to the outcome of the winning mode in the system; however, this effect is significantly weaker and cannot be quantified as the deviations from the experimental data fit remain consistently large across multiple orders of magnitude of the spontaneous emission strength $C_s$. Finally, we simulate the effects of the applied pump power to the system $\mu$, normalized by the experimental threshold current density. Increasing the pumping level of the ring laser will reduce its sensitivity to waveguide imbalances. By varying various $\mu$ values, we explore the mode switching dynamics under different laser operating regimes to be validated by future experiments.

The ability to control unidirectional mode selection and understand the switching mechanism brings us closer to realizing practical applications of controllable ring QCLs for various applications like ring-based frequency comb generation, random number generation, optical computing, and complex photonic systems on a chip.

\begin{backmatter}
\bmsection{Funding} Office of the Dean for Research, Princeton University

\bmsection{Acknowledgment} Portions of this work were presented at the International Quantum Cascade Lasers School and Workshop (IQCLSW) in 2024 under the title Optical Feedback in Quantum Cascade Ring Laser Systems. We acknowledge the Micro and Nano Fabrication Center (MNFC) of Princeton University where fabrication was carried out. 

\bmsection{Disclosures} The authors declare no conflicts of interest.

\bmsection{Data Availability Statement} Data underlying the results presented in this paper are not publicly available at this time but may be obtained from the authors upon reasonable request.

\end{backmatter}

\label{sec:refs}


\bibliographystyle{unsrt}
\bibliography{main}

\end{document}